# IDENTIFYING CRITICAL FEATURES FOR NETWORK FORENSICS INVESTIGATION PERSPECTIVES

**Ikuesan R. Adeyemi** **Shukor Abd Razak** **Nor Amira Nor Azhan**
Raikuesan2@live.utm.my shukorar@utm.my namira5@live.utm.my
Department of Computer System and Communications,
Faculty of Computer Science and Information Systems,
Universiti Teknologi Malaysia

**ABSTRACT**

Research in the field of network forensics is gradually expanding with the propensity to fully accommodate the tenacity to help in adjudicating, curbing and apprehending the exponential growth of cyber crimes. However, investigating cyber crime differs, depending on the perspective of investigation. There is therefore the need for a comprehensive model, containing relevant critical features required for a thorough investigation for each perspective, which can be adopted by investigators. This paper therefore presents the findings on the critical features for each perspective, as well as their characteristics. The paper also presents a review of existing frameworks on network forensics. Furthermore, the paper discussed an illustrative methodological process for each perspective encompassing the relevant critical features. These illustrations present a procedure for the thorough investigation in network forensics.

## 1. INTRODUCTION

Investigating how an incident occurred and who was involved, with respect to computer networks is usually referred to as network forensics. Various definition of network forensics has trailed the community of network forensics. In [2], a network forensics definition is given from the military perspective. Similarly, [3] presented a network forensics in industry paradigm. Moreover, the generally accepted description of network forensics is given in the digital forensics research workshop (DFRWS) 2001[1]. However, in this study, we defined network forensics as the study of the underlying aim, action, source and result of an attack or any incident defined to contravene organization policy, or sets of command that can result in the compromise of a system such as botnets, and malwares. The inception of system compromise or network attack is usually designed on a silent and unnoticeable process, which is often overlooked by system experts, and consequently, progress into fully-fledge attack [4]. Such techniques are developed over time, and usually, emerge within the scope of most academic syllabus [12, 13] on engineering and computer science (example include digital forensics curriculum) [15].

The academia thus plays a pivotal role [14] in the challenges rocking the digital world. Ironically, the mitigation of these challenges also resides within the confines of the academia. For effective investigation, a thorough understanding of the underlying perspective is undeniably required to answer questions relating to ‘who will be involved’, ‘what are the requirements’, ‘what resources are available and in what capacity', and 'to what end' in a decisive, wholly and reliable conclusion. The academia initiates the background knowledge required for this requirement [14, 15]. One could therefore think of the academia as the pivot upon which all aspect of network forensics is developed, without which, network forensics could stray frenzy [16].

Network forensics can be viewed from various perspectives, but the prominent ones are the military, law enforcement, civil litigation, and the network security professional. These perspectives can however, be generally classified into three [1, 37]; 'law enforcement', 'industries' and 'military'. The law enforcement perspective includes personnel in the legal technical institutions, policing system (example include first responder units), and government agencies. Industries refer to personnel in private sectors such as cyber security specialist, and organization devoted to the provision of forensic capabilities. The military perspective on the other hand refers to government military arsenal, military research institutes, as well as other military academic institution. Moreover, each of these perspectives shares similarity in varying degree of personnel, personnel qualification and responsibilities. Figure1 gives a descriptive analysis of the generic perspectives in network forensics.

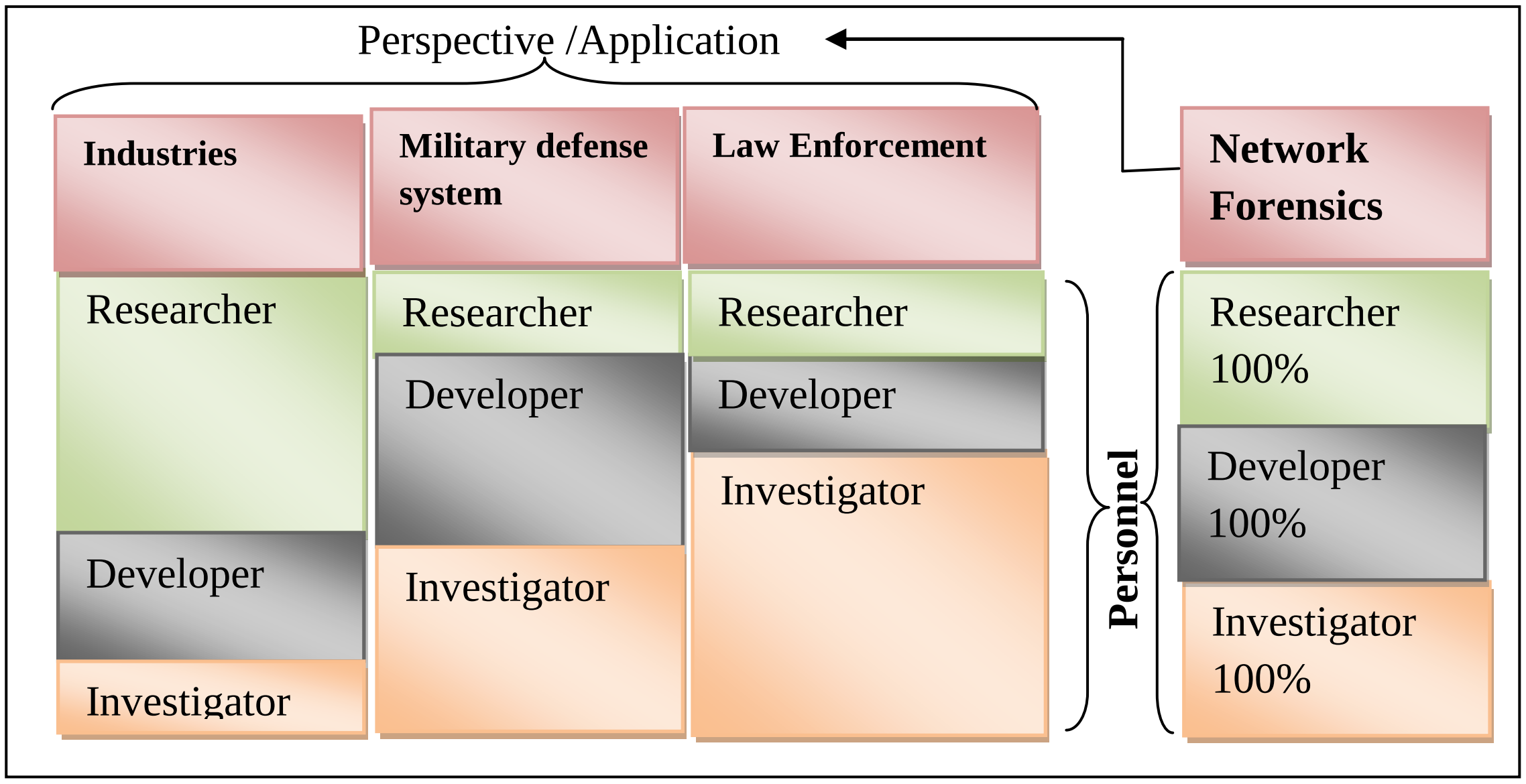

Figure 1: Perspectives of network forensics. It embodies researchers, developers, and investigators but in varying degree of scope, relevance and priority. In network forensics generic perspective, the personnel are required in almost equal proportion.

Each of these personnel: researchers, developers, and investigators shown in Figure 1, though inter-related in a loop-like relationship [1], constitute distinctly, the composition of network forensics. Researchers are personnel who undertake findings relevant to promote the existence of network forensics. Developers on the other hand are personnel who develop relevant softwares and hardware devices, needed for investigation. Investigators are personnel who engage in investigation. However, in application, each of these distinct components varies in their objective, methodology, as well as content scope. Scoping each of these perspectives to provide quantitative insight into the field of network forensics is therefore eminent, and requires urgent formulation, if network forensics discipline is to meet with its design attributes. Table 1 gives an overview of existing investigative framework for digital forensics.

Table 1. Review of existing network/digital forensic framework

| Framework/ Model | Description | Phase | | | | | | | | | | | | | | | | | | | | | | | | | | | |
|---|---|---|---|---|---|---|---|---|---|---|---|---|---|---|---|---|---|---|---|---|---|---|---|---|---|---|---|---|---|
| | | Hypothesis & Reconstruction | Planning & Notification | Authorization | Incident Closure | Digital Crime Scene | Physical Crime Scene | Awareness & Readiness | Collection & Documentation | Examination& data aggregation | Analysis & Evaluation | Reporting & Result | Preservation | Returning Evidence | Investigation initiation and Incident Response | Acquisition and monitoring | Deployment | Legal review and Presentation | Identification | Decision | Approach Strategy | Preparation and Incidence definition | Transport & Interaction | Storage & Search | Admission & Defense | Design &Implementation | Protection | Modeling and behavior prediction | Triage |
| Pollitt [36], 1995 | Cyberspace model | | | | | | | | | | x | | | | | x | | | x | | | | | | x | | | | |
| DFRWS [1] 2001 | Metal model | | | | | | | | x | x | x | | x | | | | | x | x | x | | | | | | | | | |
| Ashcroft 2001 | First responders guide | | | | | | | | x | x | x | x | | | | | | | | | | | | | | | | | |
| Reith &colleague[41] | Abstract model | | | | | | | | x | x | x | | x | x | | | | x | x | | x | x | | | | | | | |
| Carrier &Spafford[21] | Event-based scene investigation | | | | | x | x | x | | | x | | | | | | x | x | | | | | | | | | | | |
| Wei [26] | Model for information security | | | | | | | | x | x | | | | | x | | | | | | | | x | | | | | x | |
| Rowlingson[30] | Network forensics readiness | | | | | | | x | | | x | x | | x | x | | x | | | | x | x | | | | | | | |
| Beebe & Clark[39] | Hierarchical objective-based | | | | x | | | | x | | x | | | | x | | | x | | | | x | | | | | | | |
| Ciardhuain[20] | Augmented waterfall architecture | X | x | x | | | | | x | x | | | | | | x | | x | | | | | x | x | x | | | | |
| Forrester & Irwin[38] | Industrial organization model | | | | | | | x | | | x | x | x | | x | | x | | | x | | | | | | | | | |
| Rogers[34] | Field triage process model | | x | | | | | | | x | x | | | | | x | | | | | | | | | | | | | x |
| Popovsky[4] | Network forensics readiness | | | | | | | | | | | | | | | | | | | | | | | | | | | | |
| Angelopoulou[29] | ID theft Investigation framework | x | | | | | | | | | x | | | | | x | | | | | | | | | | | | | |
| Ray & colleague[45] | Domain-specific | | | | | | | | | x | x | | | | | | | | x | | | | | | | x | | | |
| Selamat & colleagues[32] | Investigation framework mapping | | x | | | | | | | x | x | x | x | x | | | | | x | | | | x | | | | | | |
| Peruma [44] | Country-based investigation process | | x | x | | | x | | | | x | | | | | | | x | x | | | | x | x | x | | | | |
| Shakeel & colleague[7] | Law enforcement framework | x | | | | | | | | | x | x | x | | | x | | | | | | x | | | | | | | |
| Pilli, & colleagues[28] | Generic framework | | | x | x | x | | | x | x | x | | x | | | | | | | | | x | | | | | x | | |
| Hunton [40] | Cybercrime investigation | | x | | | | | | | | x | | x | | x | | | | | x | | | | | | | | x | |
| Yussof &colleagues[42] | Common phase investigation model | | | | | | | | | | x | | x | x | x | x | | x | | | | | | x | | | | | |
| Agarwal & colleagues[46] | Systematic investigation | | | | | | x | | x | x | x | x | x | | | | | x | | | | x | | x | | | | | |
| Ademu, & | Activity-based | x | x | | | | | | x | | | | | | | | | x | | | | x | x | | | | | | |
| colleagues[47] | | | | | | | | | | | | | | | | | | | | | | | | | | | | | |
| Ma, & colleagues [43] | Data Fusion-based | | | | | x | x | | | x | | | | | | | | x | | | | x | | | | | | | |

Moreover, various models, and frameworks have been developed to provide insight into network forensics perspective as shown in Table 1. Though myriads of frameworks from different perspectives have trailed the community of network forensic, yet, there is no one framework that addresses the cogent features for military perspective, law enforcement perspective, and industrial perspective distinctively. Thus, this paper detailed exclusively, the critical features required for thorough network forensics investigation from law enforcement, military, and industries perspective. The rest of the paper is as follows; section 2 detailed existing frameworks and models for network forensics perspectives. Section 3 elucidate on the analysis of network forensics perspectives cueing from the various personnel. In Section 4, we present our illustrative methodological models for network forensics perspectives. Conclusion is given in section 5.

## 2. EXISTING NETWORK FORENSICS PERSPECTIVE FRAMEWORK

In [1], the first step on network forensics framework, relevant lexicon and research needs is presented. Academic researchers, military warfare, critical infrastructure protection and civil litigation paradigm were identified as the nucleus of network forensics. [2] discussed on the challenges militating against network forensics in military network environments. They identified information system of military organization as the primary victim of attack. Consequently, network forensics (in military investigation process paradigm), is described as the arsenal that provides a conclusive description of all cyber attack scenes with intent to restore critical information infrastructure, as well as to strengthen the confidence for investigative process. However, the use of network forensics simulation tools in military cyber warfare depends on specific requirement and desired aim of the organization [5,6]. The military perspective of network forensics is usually targeted at a near-real-time investigation process [8], thus, network forensics in this paradigm primarily includes the need for physical location detection and a behavior-based algorithm research, to reduce the level of cyber anonymity [5]. [6] further illustrated that military environment suffers most of the cyber attacks on critical infrastructures.

[7] proposed a 3-phased law enforcement investigation framework from law enforcement perspective. They elucidated a review of the cyber law of the “Republic of Maldives”. Similarly, [6] researched on threat mitigation for cyber investigation. In law enforcement paradigm however, traditional crime solvability is not necessarily applicable to cyber crime investigation [6], but could be applicable to threat elimination through security hardening, and crime prosecution [18]. Regardless of the level of technological improvement, investigation is human-centric (criminals, tool developers, researchers, prosecutors, investigators, and victims are human); hence a need for awareness maintenance [17] and training [4, 9] cannot be overemphasized. Furthermore, [10] expostulated that an efficient law-enforcement investigation process is one, which can facilitate relevance from contextualizing any cyber crime into a behavioral pattern, as well as quantifying the network technology for quick examination. Moreover, in [11] an extended cybercrime investigation model, for efficient cyber investigative practice in law enforcement community was proposed. In [38], a 5-phased industrial paradigm of investigation is presented. The phases include readiness, deployment, securing physical scene, securing digital scene and review phase. The readiness phase is the bedrock upon which investigation is vetted in conformance with stated organizational policy. At-scene investigative model in developed in [34]. Furthermore, timeliness in investigation was considered essentially important, through the introduction of investigation triage (a

medical terminology for prioritization) and chronology timeline. An overview of existing frameworks in presented in Table 1. Additionally, Table 2 gives a substantive synopsis of the perspective in network forensics, while Table 3, gives an elucidatory description of the various features constituting network forensics frameworks.

As shown in Table 2, the three perspectives of network forensics can be described distinctly with their characteristics, technicalities demand, critical focus, critical framework features and distinction.

### 2.1. Characteristics of Network Forensics Perspectives

Investigating network forensics differs in scope and objective from one perspective to the other. However, the scope and objective of an investigation usually depict its characteristic features. A brief description of the characteristics of the three identified perspectives are thus presented in this section

Table 2: Overview of Network Forensics Perspectives

| **Features** | **Network forensics perspective** | | |
|---|---|---|---|
| | ***Military*** | ***Law enforcement*** | ***Industries*** |
| ***Characteristics*** | Pro-active | Post-mortem investigation | Pro-active |
| | Reactive | Off-line investigation | Defensive |
| | Defensive | | Training and certification |
| | Near real time analysis | | Near real time analysis |
| | Target of attack | Investigate the target of attack | Target of attack, investigate target of attack |
| | Readily available resources | | |
| ***Similarities*** | Investigation: evidence identification, collection, fusion, analysis and documentation | | |
| ***Distinction*** | Usually near real time investigation | Post mortem investigation | Near real time investigation as well as post mortem investigation |
| | Heavy-tailed traffic type | Lightweight traffic type (usually) | Heavy tailed traffic type, and light weight traffic type |
| | Non-jurisdiction bound | Requires jurisdiction justification | Requires jurisdiction justification |
| | Inter-nations relationship | Civil litigation | Inter-city, and inter-nation relationship |
| | Low level of legal requirement | High dependency on legal protocol | High dependency on legal protocol |
| | 24/7 monitoring and analysis, strictly coordinated, hierarchical investigation process | Occasionally, and case specific investigation process | 24/7 monitoring and analysis |
| ***Technicalities demand*** | Up-to-date technologies, updated soft wares | Trusted soft ware, approved technological devices | Up-to-date technology, enhanced software, and self-automated applications |
| | High level of technological sophistication required, | Low level of technological sophistication required, | High level of technological sophistication required, |
| | highly skilled and experience personnel | highly experienced personnel | highly skilled and highly trained personnel |
| | Large network environment, and variety of homogenous (manufacturer) network devices | Relatively smaller network environment, and variety of heterogeneous (manufacturer) network devices | Large network environment and variety of homogenous (manufacturer) network device |
| ***Critical focus*** | Research centric operation | Investigation centric operation | Developer and training centric |
| | Administrative investigation provision | Litigation provision | Administrative investigation provision |
| ***Critical framework features*** | Hypothesis, Event reconstruction, Analysis, Awareness, Readiness, Incident response, Approach strategy, Investigation initiation, Authorization,<br>Modeling and behavior profiling, Risk assessment, protection, Analysis evaluation, Documentation, Reporting | Chain of custody, collection, event reconstruction, documentation, analysis, preservation, examination, acquisition, identification, Digital crime scene, Physical crime scene, | Documentation, analysis, preparation, Modeling and behavior prediction,<br>Risk assessment, protection,<br>Design, implementation, Reporting<br>Deployment, examination, chain of custody |

### 2.1.1. The Military Perspective

Network forensic investigation in the military looks beyond reactive and tactical cyber defense, to a proactive strategic cyber investigation. Military leaders have therefore begun the process of cyber investigation policy amongst which is the international military deterrence, the establishment of a Distance Early Warning Line (DEWL), and the capability to select from range of investigative arsenal [48]. As shown in Table 2, the military perspective of network forensic investigation includes;

- Proactive investigation: this type of investigation process involves the integration of expertise (expert hackers, script kiddies), motivation (financial gain, selfish aggrandizement, political achievement, personal/corporate/national vendetta, destruction), and attack vector [49, 51] of network event analysis procedure into modus operandi prediction models. Proactive investigations therefore tend to predict an event before its full incubation, by studying the underlying network traffic pattern, and intelligent correlation. This is essentially relevant for military investigation as it covers both near real time investigations, as well as ensure the readiness of resources. Additionally, such investigative paradigms are built upon the backdrop that most successful attack on military networks are heavily sponsored and could cause unredeemable catastrophic damage if successful.
- Reactive and defensive investigation: defensive investigation[53] involves identifying network vulnerabilities, and implementing necessary remedy[52] to forestall the exploitation of such loophole. Such investigation covers wide range of information security management system, and healthy network defense practice. It also involves preventing further incidence occurrence through traffic filtering and network isolation of infected host [52]. On the other hand, a reactive investigation involves investigating network device and traffic with the aim of responding to breaches, either directly, or counteractively against the intrusion source. Such investigation is defined with accuracy in identifying intrusion source, environment and underlying circumstance, as well as detail logistical information; which are reliant on the level of reliance preparedness, situational awareness, and technical expertise[14, 49]. The DOD1998 Solar Sunrise[49] is an example of such. Attacks such as the Moonlight Maze, Brazilian Power outage, and Titan Rain explicated in [54, 55] are fractions of the myriad range of threats/attacks at national infrastructure, military included.

### 2.1.2 The Law Enforcement Perspective

This perspective of investigation is carried out after an incident has occurred; a post-mortem scavenging process of network device and network related artifacts, to uncover facts substantial enough for criminal prosecution. Law enforcement investigation[56] can also include the military but for the sake of this research, we refer to law enforcement as government agencies saddled with the judicial responsibility of investigating cyber related incident, so as to provide evidence otherwise termed hidden or lost, for cyber crime related cases. Therefore, the primary responsibility of this perspective is criminal apprehension. Moreover, deterrence becomes the consequence of the investigation. Being a post-mortem investigation, it is usually an off-line or passive network evidence collection, identification, analysis, documentation, and presentation of evidence contravening stipulated law, to court of competent jurisdiction. Additionally, it exhibits reasonable expectation of prejudice[57] (a real, substantial and convincing grounds for investigation must exist before the commencement of investigation).

### 2.1.3 Industries Perspective

This perspective of investigation is relatively similar to that of military in areas of proactive, defensive investigation. As like military, it can also be the target of an attack. However more unique with this perspective is the training and certification capacity it also provides. Competent forensic investigators are usually forged from this perspective, before they are deployed or employed in other perspectives. The industries can also be described as an outsourcing unit for investigators, especially to law enforcement agencies.

## 2.2 Distinction in Network Forensics Perspective

The unique features that constitute network forensics for each of the perspective are presented in this section.

### 2.2.1. Military perspective

As identifies in table 2, network forensics in the military perspective is characterized by a stochastic heavy-tailed probability distribution (in [58], Fischer, and Fowler identified FTP transfer, page request, page reading time, session duration, session size, TCP connection, inter-arrival time of packet; to exhibit heavy tail distribution), which is due to real time or near real time analysis. Hence, most forensic tools developed in this perspective are heavy tail inclined. Moreover, investigation in this perspective functions autonomously of jurisdictional boundaries, and does not require any special court order to react, defend, or initiate investigation. However, monitoring, and event analysis, is strictly coordinated and usually follow a hierarchical model of clearance level evaluation such as the Bella Padula model [59].

### 2.2.2 Law Enforcement Perspective

This perspective is case specific, and adheres strictly to legal regulation. Since it has to do with evidence integrity, and admissibility in court of competent jurisdiction, law enforcement perspective requires jurisdictional justification, approved search and seizure warrant, well documented chain of custody note (see table 3), and transparent investigative process. The strict observance of legal protocol is a cardinal part of law investigation.

### 2.2.3 Industries Perspective

Investigation in this perspective is derives its uniqueness from both the military and the law enforcement. It is relatively similar to the military as well as law enforcement perspective in term of investigation type (near real time or offline), autonomous investigative process, inter-city and inter-national boundaries, and . However, this perspective can grow beyond the capacity of any military or law enforcement or both. Thus, an industrial perspective can be more complex to describe but maintains certain unique features nonetheless.

The various technicalities demand for each perspectives as well as the critical focus are presented in Table 2. However, the critical framework features (see Table 3) are further discussed in the proceeding section.

## 3. FEATURES OF NETWORK FORENSICS PERSPECTIVE

The criticality of network forensic feature depends largely on the perspective, size, topology, and expertise of the investigator. The choice of feature to include in investigation, also describe the expected thoroughness of the investigation. In this section, we present the features that are critical for network forensic investigation for the three perspectives.

Moreover, a concise descriptive definition of features used in network forensics is presented in Table 3. These features are derived from existing framework on digital forensics investigation. The term 'Ff' is an abbreviation for framework features. As noted in Table 3, some features are essential for all perspective irrespective of the crime scene involved. However, some are unique to certain perspective, which when included into the investigative process of other perspectives could result in higher overhead running cost (in term of resources and efficiency) and redundancy of service.

Table 3: Framework Feature Description

| ***Framework feature*** | ***Description*** | ***Perspective critical to*** |
|---|---|---|
| ***Ff1***-Chain of custody | Chain of custody is a concept usually a written material that contains all processes carried out before, during and after an investigation on 'what was done', 'why it was done', 'who did it', and 'when it was done' [17, 19], a documentation proofing the integrity of evidence [18]. | All network forensics perspectives |
| ***Ff 2***-Hypothesis | It is a supposition or proposition put forward by an investigator, as an explanation to an occurrence, to initiate an investigation based on evidence examination [20, 21]. Hypothesis usually followed the SMART (specificity, measurability, attainability, realistic, and timeliness) ideology consideration. | Industries, Military |
| ***Ff 3***-Reconstruction | Event reconstruction is the process of reconstructing the sequence of network traffic [22], from captured traffic accumulated, and or network device logs and other related devices, for establishing an occurrence and its supporting artifacts.[21,23]. The use of NFAT in the network forensics community today, has made this process easier, but still requires more consolidated and efficient technique, for undisputable evidence analysis process. | All network forensics perspectives |
| ***Ff 4***-Authorization | Investigation authorization involves the granting of legal permission to the effect of commencing investigation process. This could also involve the acquisition of a search and/or seizure warrant from a court of competent jurisdiction.[16] | Military |
| ***Ff 5***-Incident closure | This is the process of closing a particular network investigation exercise, usually after appropriate satisfactory status. It is preceded by a thorough review of the entire investigation process, well-articulated chain of custody, documentation and expert review consideration [24]. | Law enforcement |
| ***Ff 6***-Digital crime scene | Securing the digital crime scene involves the practice of strict adherence to safe digital procedure for evidence acquisition, and preservation. It describes the ethics of first responders and computer emergency response team (CERT), to digital crime scene due to fragility and volatility of network forensics evidence [25]. | Law enforcement |
| ***Ff 7***-Physical crime scene | Securing the physical crime scene involves the practice of due caution, and professionalism in safeguarding crime scene, and the use of appropriate signage. It generally describes the responsibility of first responders, and CERT [25] | Law enforcement |
| ***Ff 8***-Awareness | It is usually associated with staff training on updated knowledge in network forensics [[17]. Staffs include CERT, and organization IT staffs. | Military, Industries |
| ***Ff9***-Readiness | This is the act of being prepared for investigation at any given time. It combines section of organs of an organization for preparedness in the event of an emergency, as well as anticipated event of network intrusion breach. | Industries, military |
| ***Ff 10***-Collection | This is the process of collecting network traffic information for investigation purpose. It usually takes reasonable period, and in a pre-event-occurrence process. Due to network traffic volatility, evidence collection involves the combination of both network hardware and software composition [16, 26]. | All network forensics perspectives |
| ***Ff 11***-Documentation | This is the process of taking account of every process and activities carried out during investigation and the reason why it was done in such as manner [27]. It is the heart of investigation, and contains, strictly articulated write-up of the entire investigation procedure. Documentation also serves as expert review, examiners' note; | All network forensics perspectives |
| | source for future event investigation [17, 16, 25]. | |
| ***Ff 12***-Examination | This is the process of scavenging network traffic for clue or sample of relevant incriminating evidence. Devices to be examined include but not limited to, network devices. Examination could be static/manual process or automated process. | Law enforcement |
| ***Ff13***-Analysis | Analysis is sometimes categorized as examination. According to [27], it is the "process of interpreting extracted data, to ascertain the level of relevance or significance to ongoing investigation process". Network forensics analysis tools (NFAT)[28] are usually adopted for this phase (time framing analysis, data hiding/steganography analysis[27]) of network forensics. It is also the application of validated techniques to discovering or uncovering significant data [32] | All network forensics perspectives |
| ***Ff 14***-Evaluation | Evaluation could be prior to evidence analysis, in this case, it reviews the facts required for examination; during evidence analysis, in this case, to determine the accuracy, thorough objectivity of the investigation, as well as conformity to stated priorities; or post event analysis, which involves the review of resultant artifacts, to proposed hypothesis, or other related undisputable facts. It is the process of deciding whether to accept or reject facts uncovered [32] | All network forensics perspectives |
| ***Ff 15***-Preservation | This is the acts as well as the process of ensuring that the state of a particular network traffic evidence is not altered before, during or post event analysis. This is crucial to investigations requiring further analysis or other independent investigation. Preservation is a major factor for evidence admissibility in civil litigation. | Law enforcement |
| ***Ff 16***-Returning of evidence | This is the process of ensuring that all evidence collected during investigation are safely return to its supposed owner, and in the same or almost the same condition at the seizure and acquisition state. | Law enforcement |
| ***Ff 17***-Investigation initiation | This includes history from previously investigated cases. Initial investigation is the process of gathering relevant artifacts about a particular investigation process, building a predefined network traffic behavior database to ease (with respect to time, resources, and methodology) in investigation process. it marks the beginning or call for investigation | Industries, Law enforcement |
| ***Ff 18***-Acquisition | This is the process of gathering or gaining possession [31] to network traffic artifacts for or during investigation. | Law enforcement |
| ***Ff 19***-Deployment | This involves putting in place respective forensics measure for proper conduct of investigation. According to [21, 30], deployment can be initiated after thorough evaluation of inputs from network security agent. | Industries, Military |
| ***Ff 20***-Presentation | This is the act of presenting authoritatively, the investigated facts, to relevant constituted authority. It is usually carried out as the last stage of network forensics investigation phases. | Law enforcement |
| ***Ff 21***-Identification | This is the process of pinpointing or locating relevant network forensics evidence from database of network traffic or from stream of traffic flow. An adequate and precise identification process goes a long way in influencing the amount of resources, the duration of investigation, and the weight of the evidence. | All network forensics perspectives |
| ***Ff 22***-Decision | This is the process of attributing certain parameters, artifacts of evidence and concluding on the result of the analysis from the investigation. This stage is the most critical phase of investigation, and it requires a thorough review of the entire process, expert counsel, and experience where necessary. | All network forensics perspectives |
| ***Ff 23***-Approach | This describes the designed process adopted for the investigation flow. A choice of which phase to carry out, | Military |
| strategy | and in what sequence, and with what resources and in what manner. Approach strategy is a decision making process which usually involves expert input, in line with organization policies. | |
| ***Ff 24*-**Preparation | This is the process of organizing the necessary network forensics requirement and process for investigation. It also involves the timely dissemination of investigation procedure and schedules to affected parties.[32] | Industries, military |
| ***Ff 25*-** Transportation | This is the process of moving collected network related evidences from one place to another (usually a network forensics laboratory) through a secure channel and procedure, in a well-documented order, and duly appended in chain of custody. | Law enforcement |
| ***Ff 26*-** Interaction | This is the process of communicating relevant investigation process or result to constituted authority [32], with the view of sharing idea, developing better evidence decision process, and or demonstrates the level of investigation success. | Military |
| ***Ff 27*-** Storage | This is the process of storing network related artifacts. This process usually involves well-established storage and retrieval mechanism, with a proper write/read blocker. | All network forensics perspectives |
| ***Ff 28*-** Search and seizure | This is usually attributed to legal warrant obtained for the commencement of investigation. It involves the permission from constituted legal authority to carry out search on the victim or suspect system for relevant or incriminating evidence, and when necessary, seize the evidence source for thorough investigation [32]. | Law enforcement |
| ***Ff 29*-** Admission | This involves the taking-in of a particular network traffic data as part of the sources of network forensics evidence. Admitting evidence in network investigation process also involve the process of acknowledging and accepting an evidence as an authentic, and genuine. | Law enforcement |
| ***Ff 30*-** Defense | This is the process of preventing alteration of network evidence, in order to maintain its integrity. Evidence defense also encompass the act of ensuring that a thorough explanatory analysis is provided to backup supposition and result of the analysis. | Law enforcement, Military |
| ***Ff 31*-** Design and implementation | This is the process of establishing a workable network forensics investigation pattern and methodology for a particular investigation process. it usually stern from organization policies, and investigator's experience from previously investigated scenes | Military, industries |
| ***Ff 32*-** Protection | Is the process of preventing network traffic alteration before, during or after investigation. It is also the practice of ensuring integrity and validity of evidence for future use, or reference [32]. | All network forensics perspectives |
| ***Ff 33*-** Risk assessment | This is the act as well as process of taking into consideration the various factors involves for network forensics investigation so as to understand the risk at stake before initiating an investigation. Furthermore, risk assessment is the critical examination of organizations assets to identify assets that can justify legal redress when deliberately compromised [33]. | Industry |
| ***Ff 34*-** Modeling and behavior prediction | This process involves the mathematical or analytical procedure for forecasting the possibilities of event occurrence, to accelerate investigator's decision–making process [34]. Network forensics modeling and behavior prediction is a complex process that, when properly carried out, can improve the efficiency of network analysis. | Military, industry |
| ***Ff 35*-** Data aggregation | Data aggregation in network forensics is the process of clustering independent, but similar featured network traffic. This is executed in a coherent and methodological procedure to speed up investigating time. The process of significant features identification for data aggregation is given in [35]. | All network forensics perspective. |
| ***Ff 36* -** Triage | Network forensics triage is the process of sorting and prioritizing methodology, and investigative process, in order to increase the overall efficiency of the analysis, evaluation and decision making process. In [34], a field triage model was defined to catalyze the period required for investigation. | Law enforcement |

## 4. ILLUSTRATION OF PERSPECTIVE METHODOLOGY

In Table 2, a detailed overview of the characteristics, similarities, technicalities, and focus of each of the perspectives are described. In this section, we present the proposed models for each of the perspectives.

Table 4: Critical Features for Network Forensics perspectives

<table>
<tr><th>Perspective</th><th>Critical features</th><th>Investigation process</th></tr>
<tr><td>Military</td><td>Ff 1 + Ff 2 + Ff 3 + Ff 4 + Ff 8 + Ff 9+ Ff 10 + Ff 11 + Ff 13 + Ff 14 + Ff 19 + Ff 21 + Ff 22 + Ff 23 + Ff 24 + Ff 26 + Ff 27 + Ff 30 + Ff 31 + Ff 32 + Ff 34 + Ff 35+ Ff 36</td><td rowspan="3">• Thorough understanding of the unique investigation scenarios in each perspective<br>• Selection of features using a sequential methodology, such as appropriate for investigation development life cycle for each perspective<br>• Acceptable definition and scope of each features/phases based on organization policy, electronic/multimedia/communication Acts of the country, international laws</td></tr>
<tr><td>Law enforcement</td><td>Ff 1 + Ff 3 + Ff 5 + Ff 6 + Ff 7 + Ff 10 + Ff 11 + Ff 12 + Ff 13 + Ff 14 + Ff 15 + Ff 16 + Ff 17 + Ff 18 + Ff 20 + Ff 21 + Ff 22 + Ff 25 + Ff 27 + Ff 28 + Ff 29 + Ff 30 + Ff 32 + Ff 35 + Ff 36</td></tr>
<tr><td>Industries</td><td>Ff 1 + Ff 2 + Ff 3 + Ff 8 + Ff 9 + Ff 10 + Ff 11 + Ff 13 + Ff 14 + Ff 17 + Ff 19 + Ff 21 + Ff 22 + Ff 24 + Ff 27 + Ff 31 + Ff 32 + Ff 33 + Ff 34 + Ff 35+ Ff 36</td></tr>
</table>

### *4.1 Military perspective illustration*

The military perspective highlighted in Table 2 reveals that network forensics in this paradigm requires an updated real-time validation. However, before any action can be taken from a real-time analysis, thorough investigation must be presented in manner consistent with the military combative methodology. Hence, in table 4, detailed critical feature for in-depth investigation is presented. Features such as Ff1, Ff11, and Ff8 are primarily critical for decision defense in military paradigm of investigation; hence, they cut across the entire phases of investigation procedure presented in figure 2.

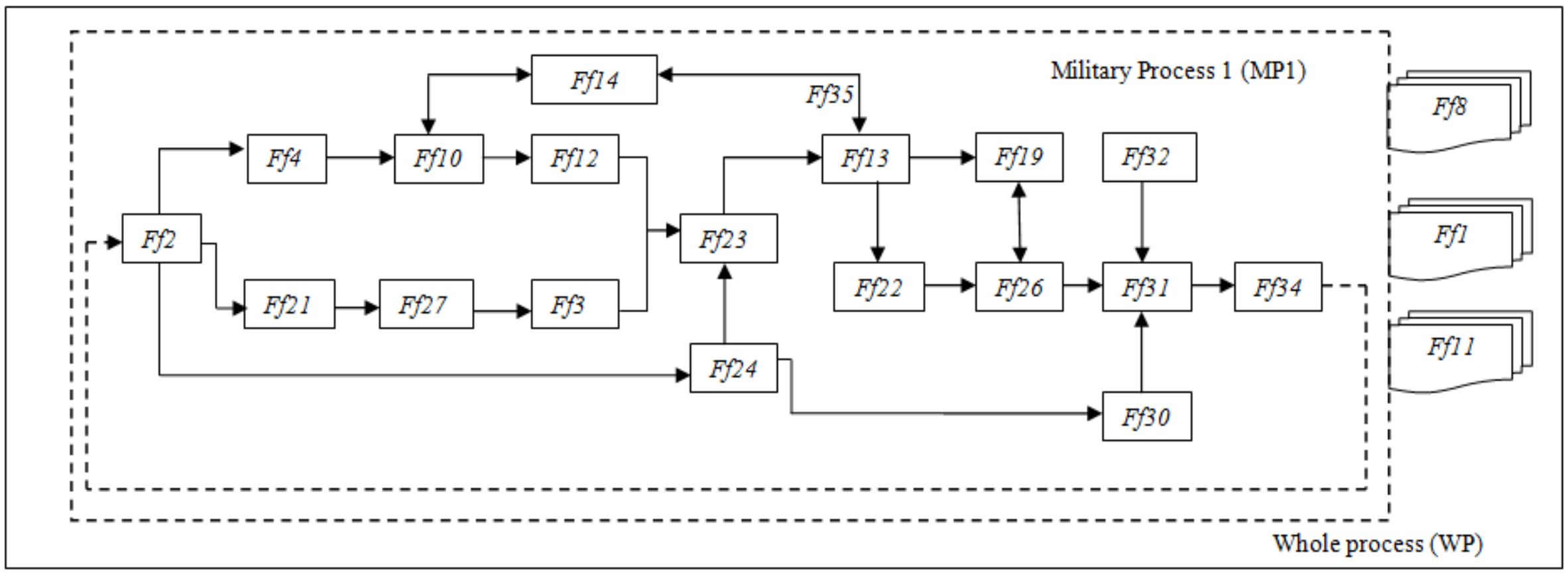

Figure 2: Network forensics military investigative perspective illustration.

Figure 2 is a 19-phase (with additional 3-phase attached to each phases) investigation illustration for network forensics. The MP1 procedure can be further translated into the following sequential procedure.

- Ff2,+ Ff4 + Ff10[(Ff14+ Ff35)], +(Ff12+Ff23)], +Ff13, +[Ff19+Ff22], +Ff26,+ Ff31(Ff 32+ Ff 30) + Ff34 + Ff …(1)
- Ff2 + Ff21 + Ff27 + Ff3 + Ff23 + Ff13 (Ff35 + Ff14)+ [Ff19 + Ff22] + Ff26 + Ff31(Ff32 + Ff30) + Ff34 + Ff2 …(2)
- Ff2 + Ff24 +[ Ff30 +( Ff23 + Ff13 +( Ff19 + Ff22) + Ff26 + Ff32] + Ff31 + Ff34 + Ff2 …(3)

In contrast to other existing model, this illustration adopts a recursive iteration procedure that can help to reduce possibilities of human error, as well as overlooked facts. Additionally, it reduces investigation overhead accumulated due to features clustered phases.

### *4.2 Law Enforcement Perspective Illustration*

In Table 2, law enforcement paradigm in network forensics investigation process is characterized by post event occurrence. Thus, an in-depth postmortem in scavenging network devices and stored databases is required for a network forensics investigation. Moreover, investigation procedure differs from one crime scene to another, and usually depends on the discretion of the investigator. Hence, Figure 3 depicts an illustrative methodology for network forensic investigation.

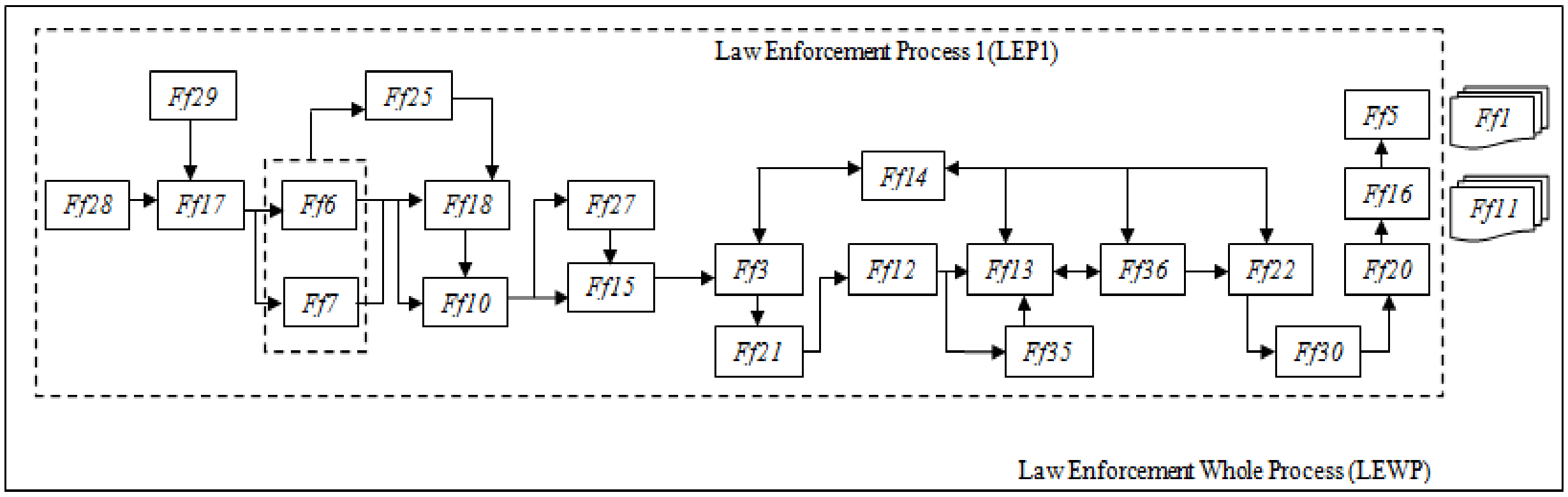

Figure 3: A Network forensics law enforcement perspective model. Features outside the LEP1, are carried out simultaneously with each of the phases in LEP1. This is required for effective documentation as well as to improve the quality of expert witness.

This illustration involves a 23-phase investigative procedure, which are further translated into the following:

- Ff28+ Ff17+[ Ff6+ Ff7+( Ff8+ Ff10)]+ Ff25+ [Ff27+ Ff15]+Ff3+Ff21 + Ff12 +( Ff13+ Ff35)+ Ff36+ Ff14+ Ff22+ Ff30+ Ff20+ Ff16+ Ff5 …(4)
- Ff29+ Ff17+[ Ff6+ Ff7+( Ff8+ Ff10)]+ Ff25+ [Ff27+ Ff15]+Ff3+Ff21 + Ff12 +( Ff13+ Ff35)+ Ff36+ Ff14+Ff22+ Ff30+ Ff20+ Ff16+ Ff5 …(5)

Irrespective of the procedure of choice, this example can be seen as a non-recursive investigative process. The translated procedure in ‘1’ and ‘2’ above terminates on same feature (Ff36+ Ff14+Ff22+ Ff30+ Ff20+ Ff16+ Ff5), further indicating that the law enforcement paradigm of investigation can be termed a project-like investigation.

### *4.3 Industry perspective Illustration*

Figure 4, depicts an investigative illustration for industries. However, depending on the organizational management policy, some features could be skipped. It involves a 18-phase (with additional two for each phases) forensics procedure, which can be translated as

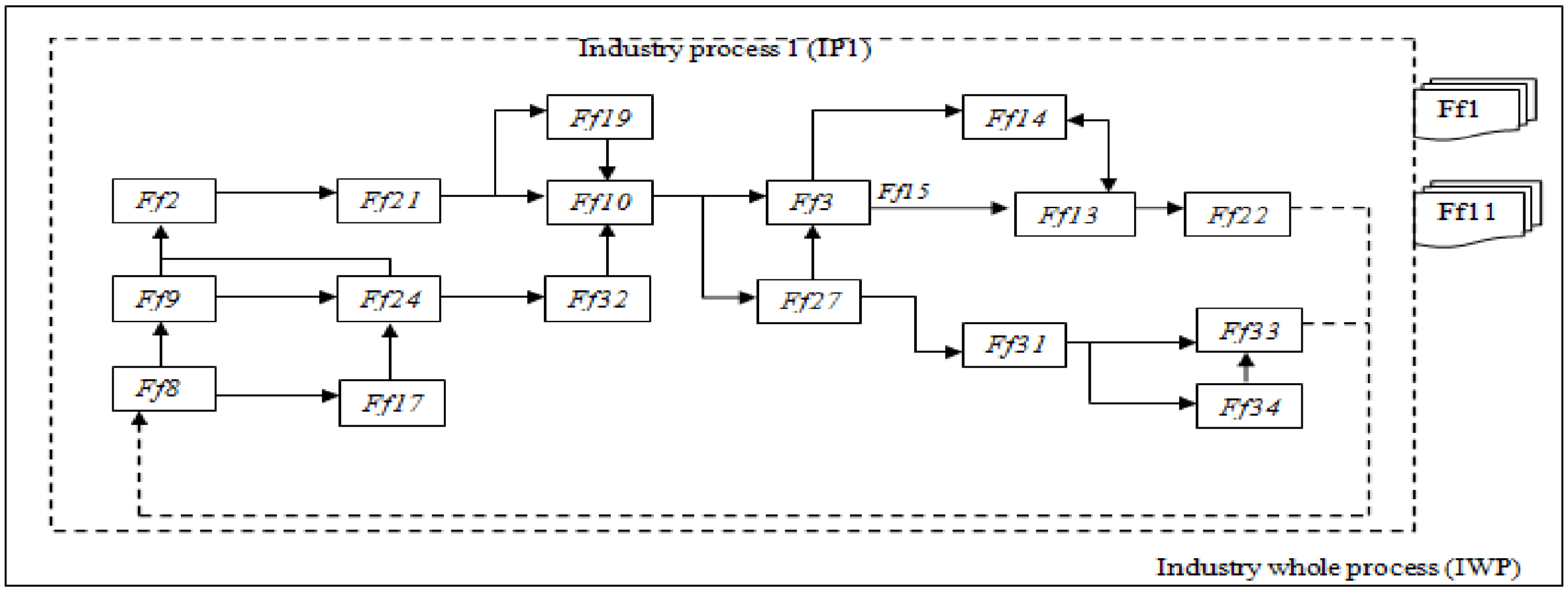

Figure 4: An industry perspective illustration of network forensics investigation

- Ff8+ Ff17+ Ff24+[ Ff32+( Ff2+ Ff21+ Ff19)]+ Ff10(Ff32)+[ Ff3(Ff14+ Ff15)+ Ff27]+ Ff13+ Ff22 …(6)
- Ff8+ Ff17+ Ff24+ Ff2+ Ff21+[ Ff10(Ff19)] + Ff10(Ff32)+ Ff27+ Ff31+[ Ff34(Ff33)]+ Ff8 …(7)
- Ff8+ Ff9+ Ff2+ Ff21+[ Ff19+Ff10(Ff32)]+ Ff27+ Ff31+[ Ff34(Ff33)]+ Ff8 …(8)
- Ff8+ Ff9+ Ff2+ Ff21+[ Ff19+ Ff10(Ff32)]+ [ Ff3(Ff14+ Ff15)+ Ff27]+ Ff13+ Ff22 ...(9)

Each of the above translation distinctly forms a pattern thorough enough for investigation. However, the combination of the features defined in IWP can yield a more thorough investigation result.

## 5. DISCUSSION

Each of the illustrations can be further translated into the highlighted dimension in equation 1 to 9. In Figure 2, 'IWP' comprises the 'IP' combined with the Ff1, Ff11 and Ff8. The Ff8 feature is considered critical due to the need for constant awareness of latest attack pattern, evolutionary network malwares, and up-to-dated network defense arsenal. Ff1 feature is a critical feature for all network forensics procedure as its forms the reservoir for knowledge on evidence detail at every event and process carried on before, during, and after investigation. Similarly, Ff11, serves as the knowledge deposit for event procedure, as well as resource for proper investigation evaluation, and expert witness note. The integration of critical features Ff1, Ff11 and Ff8 into the translated procedures in equations 1, 2 and 3, provide investigators vintage view of the investigation. Additionally, in Figure 3 'LEWP' represents the entire investigation procedure for the model. Ff1 and Ff11 features are integrated in every step in the model. Moreover, in Figure 4 'IWP' integrates Ff1, and Ff11 into each step in the investigative model.

With this illustration, network forensics can thoroughly scavenge network devices in a methodological procedure. The GCFIM model proposed in [42] by Yussof, Ismail and Hassan, (2011), identified presentation, preservation, planning, identification, examination, collection and analysis with value of 7, 4, 3, 6, 5, 6, 7 respectively, as the common features for investigation from a survey of 14 frameworks. However, they failed to identify any specific perspective of application of their 5-phased framework. Moreover, with description and analysis from this research a thorough analysis and choice of feature deemed critical to the relevant investigation process can be selected/adopted. Furthermore, a logical sequential and or iterative methodological principle can be applied.

## 6. CONCLUSION

In this paper, we discussed the existing network forensics frameworks. Special attention was directed towards the three major perspectives (as identified in most research works, particularly, [1], & [37]) of network forensics. Furthermore, we identified the critical features required for thorough investigation, and we synthesize extensively, the various perspective of network forensics. Based on the identified features, we demonstrated illustrative procedures that can be used to integrate these critical features for each perspective.

We hope to conduct extensive experimental process on these illustrations, in our research on network forensics analysis and experimental works on insider misuse prevention. Additionally, we hope to fully integrate these illustrations into an automated investigative process useful to the cyber policing community, as well as research community, thus limiting investigators prerogative in investigation process.